\documentclass{ws-p9-75x6-50}
\input psfig.tex
\begin{document}

\title{Generation and Acceleration of Jets from Effective Boundary Layer around Black Hole}

\author{Indranil Chattopadhyay and Sandip K. Chakrabarti}

\address{S.N. Bose National Centre for Basic Sciences, JD-Block, Salt Lake, Calcutta
700098\\E-mail: indra@boson.bose.res.in and  chakraba@boson.bose.res.in}

\maketitle

\abstracts{We propose the processes of generation and radiative acceleration of jets from the
centrifugal pressure dominated boundary layer around a black hole.}

\noindent Proceedings of the 9th Marcel Grossman Meeting in Rome (Ed. R. Ruffini)

The inner boundary condition at the horizon for an inflow is that the 
radial flow velocity should be equal to the velocity of light. At an intermediate 
distance, centrifugal force becomes comparable to the gravity and matter is slowed down
making it hot and puffed up like a tori. This is known as 
the centrifugal barrier supported boundary layer or CENBOL\cite{ct95,dc99}.
This CENBOL acts as the effective boundary layer around black holes.
Chakrabarti and Titarchuk\cite{ct95} calculated the spectrum emitted
from the CENBOL and showed that the CENBOL is the elusive `Compton cloud'
which gives rise to the power law hard tail. They
also showed that if the accretion rate (${\dot M}_d$)
of the Keplerian component (the generator of soft photons) 
of the disc is greater than $\sim 0.4 {\dot M}_{Edd}$, then the 
emitted spectrum will be without the hard tail due to thermal 
Comptonization and the black hole would be in the soft state. 
Subsequently it was realized\cite{ijprev,chak99,dc99}
that the CENBOL is not just a source of Comptonization but also acts as 
the source of jets and outflows. A part of the in falling matter is 
deflected by this effective boundary layer or CENBOl along the axis of symmetry
to produce jets and outflows in at least some of the observed cases.
This computed mass loss rates\cite{chak99,dc99} generally agree with 
the rates obtained through fully time dependent numerical simulations\cite{mlc94}.
Recent reports on galactic black hole candidate GRS 1915+105\cite{f2000}
and extragalactic black hole M87\cite{jbl99} clearly show evidence that
the region where jets originate and the region which emits Comptonized photons
are the same. This vindicates our stand that the advective flow solution
proposed more than ten years ago and which included accretion and winds 
interfaced by a CENBOL could be the correct paradigm.

Since CENBOL produces funnel wall as in a thick disk, and since matter 
and hard radiation can be thought of being emitted by the CENBOL surface,
the effect of momentum deposited by the hot photons on outflows could be interesting to study.
The effect of radiative momentum deposition on astrophysical outflows 
around compact objects were studied using early disc models by many workers
(e.g. Vincent Icke\cite{vi80} and Abramowicz and Piran\cite{ap80}).
Chattopadhyay and Chakrabarti\cite{1cc00,2cc00}, revisited the problem
using the radiation from advective discs\cite{ct95}. 
In this region, the subsonic outflowing matter is
continuously bombarded by hot photons and hence apart from
usual thermal acceleration, radiative acceleration should be important
too. Thus the toroidal region or the CENBOL of accretion disc not only
emits high energy radiation, but its typical shape helps the
radiation to be focussed on the axis of symmetry. 
In our calculation of radiative momentum deposition force (RAMOD) we do not consider 
the attenuation of the radiation field due to interaction with matter. With attenuation taken into account,
optically thick flow would be difficult to accelerate this way. The two components of radiative momentum 
deposition force (RAMOD) have been calculated\cite{ccheid00} and efforts are on
include these forces in the analytical and numerical study.
Our assumption of optically thin flow (i.e., $\tau <1$) puts upper limit on the mass outflow rate
and the mass of the central object\cite{2cc00,ccheid00}:
$$ 
\frac{{\dot M}^6}{{M_B}^5}<1.7{\times}10^{-70} {\rm gm\ sec^{-6}}
$$ 
or, 
$$ 
{{\dot m}_{10}}^6m_{10}<6.1{\times}10^{-8},
$$ 
where, ${\dot m}_{10}$ is the mass outflow rate ($\dot M$)
in the units of Eddington rate for a $10 M_{\odot}$ black hole and $m_{10}$ is the black hole
mass ($M_B$) in units of $10 M_{\odot}$. Increasing ${\dot M}_{acc}$ increases disk luminosity.
This in turn increases the  magnitude of RAMOD and the outflow rate ${\dot M}_{out}$. This
increases the density $\rho$ of the outflow thereby increasing $\tau$ and causing an
attenuation of radiation by a factor of $e^{-{\tau}}$.
Using the RAMOD we find the terminal speed of outflows achieved are about hundred percent
higher than that achieved by purely thermally driven outflows. Of course, radiation drag force
are seen to reduce the acceleration by some amount. But the drag is not very effective 
and we are unable to reach close to velocity of light by hydrodynamical 
processes alone. Thus our conclusion 
is that RAMOD does accelerate the outflow and higher RAMOD results in generally higher terminal speed.
RAMOD also increases the energy of the flow often freeing bound matter (i.e.,
matter with negative energy) from the accretion disk.

This work is partly supported by DST project Analytical and Numerical
Studies of Astrophysical Flows Around Compact Objects.


\begin{thebibliography}{99}
\bibitem{ct95} Chakrabarti S.K. and Titarchuk, L.G.  {\it Astrophys. \ J.}\ {\bf 455}, 623 (1995)
\bibitem{ijprev} S.K. Chakrabarti, {\it Ind. J. Phys.}, 72B, 183 (1998) astro-ph/9803227.
\bibitem{chak99} Chakrabarti, S. K. {\it Astron. Astrophys.}, {\bf 351} 185 (1999)
\bibitem{dc99}  Das T.K. and  Chakrabarti S.K. {\it Class.\ Quant.\ Grav.}\ {\bf 16}, 3879 (1999)
\bibitem{mlc94} Molteni, D. Lanzafame, G. and Chakrabarti, S. K {\it Astrophys. J.}, {\bf 425} 161 (1994)
\bibitem{f2000} Fender, R. P. {\it astro-ph/9911176} (2000)
\bibitem{jbl99} Junor, W. Biretta, J. and Livio, M. {\it Nature}, {\bf 401} 891 (1999)
\bibitem{vi80} Icke, V. {\it Astron. J.}, {\bf 85}(3) 329  (1980).
\bibitem{ap80} Abramowicz, M. A. and Piran, T. {it Astrophys. Journ.} {\bf 241} L7 (1980).
\bibitem{1cc00}  Chattopadhyay, I. and Chakrabarti, S. K.  {\it Int. Journ. Mod. Phys. D}, 
 {\bf 9}(1) 57 (2000).
\bibitem{2cc00} Chattopadhyay, I. and Chakrabarti, S. K.  {\it Int. Journ. Mod. Phys. D},
 {\it in press}
\bibitem{ccheid00} Chattopadhyay, I. and Chakrabarti, S. K. in  {\em Proc.  of High Energy Gamma-Ray Astronomy}
eds. F.A. Aharonian and H. V\"olk (in press)
\end{thebibliography}
\end{document}